\documentclass[aps,prc,twocolumn,showpacs,superscriptaddress,floatfix]{revtex4-1}
\usepackage{bm}
\usepackage{graphicx,amssymb}
\usepackage{multirow,amsmath,array,booktabs,color}

\usepackage{CJK}

\usepackage{braket}

\begin{document}

\begin{CJK*}{GBK}{song}

\title{Role of quadrupole deformation and continuum effects in the ``island of inversion'' nuclei $^{28,29,31}$F}

\author{Yu-Xuan Luo}
\affiliation{School of physics and materials science, Anhui University, Hefei 230601, P.R. China}

\author{K\'{e}vin Fossez}
\affiliation{FRIB Laboratory, Michigan State University, East Lansing, Michigan 48824, USA}
\affiliation{Physics Division, Argonne National Laboratory, Lemont, Illinois 60439, USA}

\author{Quan Liu}
\email[E-mail:]{quanliu@ahu.edu.cn}
\affiliation{School of physics and materials science, Anhui University, Hefei 230601, P.R. China}

\author{Jian-You Guo}
\affiliation{School of physics and materials science, Anhui University, Hefei 230601, P.R. China}

\date{\today}

\begin{abstract}
	\begin{description}
		\item[Background]
			The peculiar properties of nuclei in the so-called ``island of inversion'' around $Z=10$ and $N=20$ are the focus of current nuclear physics research.
			Recent studies showed that $^{28}$F has a negative-parity ground state and thus lies within the southern shore of the island of inversion,
			and $^{29}$F presents a halo structure in its ground state,
			but it is unclear which effects, such as deformation, shell evolution due to tensor forces, or couplings to the continuum, lead to this situation.
		\item[Purpose]
			We investigate the role of quadrupole deformation and continuum effects on the single-particle structure of $^{28,29,31}$F
			from a relativistic mean-field approach, and show how both phenomena can lead to a negative-parity ground state in $^{28}$F and halo structures in $^{29,31}$F.
		\item[Methods]
			We solve the Dirac equation in the complex-momentum (Berggren) representation for a potential with quadrupole deformation at the first order 
			obtained from relativistic mean-field calculations using the NL3 interaction, 
			and calculate the continuum level densities using the Green's function method.
		\item[Results]
			We extract single-particle energies and widths from the continuum level densities to construct the Nilsson diagrams of $^{28,29,31}$F in the continuum,
			and analyse the evolution of both the widths and occupation probabilities of relevant Nilsson orbitals in $^{28}$F and find that some amount of prolate deformation must be present. 
			In addition, we calculate the density distributions for bound Nilsson orbitals near the Fermi surface in $^{29,31}$F 
			and reveal that for a quadrupole deformation $0.3 \leq \beta_2 \leq 0.45$ (prolate), characteristic halo tails appear at large distances. 
		\item[Conclusions]
			Using the relativistic mean-field approach in the complex-momentum representation with the Green's function method,
			we demonstrate that in neutron-rich fluorine isotopes, while in the spherical case the $pf$ shells are already inverted and close to the neutron emission threshold,
			a small amount of quadrupole deformation can dramatically reduce the gap between positive- and negative-parity states and increase the role of continuum states, 
			ultimately leading to the negative parity in the ground state of $^{28}$F and the halo structures in $^{29,31}$F. 
	\end{description}
\end{abstract}

\maketitle

\section{Introduction}

The description of the so-called magic numbers 2, 8, 20, 28, 50, and 82 for both protons and neutrons,
as well as 126 for neutrons, in stable nuclei \cite{Haxel1949,Mayer1949} was a major achievement
for our understanding of nuclear structure close to $\beta$-stability.
Magic numbers emerge due to large energy gaps in the single-particle structure of nuclei,
which results in closed-shell nuclei with increased stability, spherical ground states and first $2^+$ excited state,
and smaller $B(E2)$ values than their neighbors on the nuclear chart.

However, later experiments showed a weakening or even disappearance of the shell effects associated with magic numbers in closed-shell nuclei away from $\beta$-stability.
Thibault \textit{et al.} \cite{Thibault1975} first reported the observation of the disappearance of the magic number ($N=20$)
based on mass measurements of the neutron-rich nuclei $^{31,32}$Na.
Later on, similar observations were made in several other neutron-rich nuclei around $N=20$ \cite{Huber1978,Detraz1979,Guillemaud1984,Motobayashi1995,Yanagisawa2003},
indicating strong deviations of the shell ordering compared to the standard shell model \cite{Caurier2005}.

The breakdown of the $N=20$ shell closure has been discussed extensively in Ref.~\cite{Warburton1990,Sorlin2008,Otsuka2020}
and is associated with the emergence of deformation \cite{Hamamoto2004,Hamamoto2009,Hamamoto2012},
the natural shell evolution due to tensor forces \cite{Otsuka2001,Otsuka2005}, and weak binding.
The rearrangement of the single-particle structure in this region \cite{Poves1987,Caurier2005,Tsunoda2020,Miyagi2020} favors a strong mixing
of the positive-parity $sd$ shells with negative-parity $fp$ shells or intruder shells,
leading to parity inversions in the ground states of several nuclei as compared to the standard shell model predictions.
Such nuclei, in which intruder configuration dominate the ground state, form the so-called island of inversion (IOI).
In recent years, with the development of experimental technology, the boundaries of the IOI have been greatly expanded.
Some nuclei, such as $^{28,29}$F \cite{Revel2020,Gaudefroy2012}, $^{28}$Ne \cite{Terry2006}, $^{29,30}$Na \cite{Tripathi2005,Tripathi2007,Seidlitz2014,Petri2015},
and $^{31,35,36}$Mg \cite{Neyens2005,Gade2007,Gade2011}, have all been experimentally confirmed as part of the IOI.

Neutron-rich fluorine isotopes lie near or on the southern shore of the IOI.
In recent years, a series of experiments had been conducted on these isotopes to explore the impact of intruder states
on their single-particle levels \cite{Elekes2004,Schiller2005,Gaudefroy2012,Christian2012,Christian2012b,Doornenbal2017,Ahn2019,Revel2020}.
In particular, the isotope $^{28}$F present an interesting situation.
It was found to be unbound with respect to neutron emission by Schiller \textit{et al.} \cite{Schiller2005}
from its relatively low upper limit for production in the reaction products of a secondary beam of $^{29}$Ne.
Later, using invariant mass spectroscopy,
Christian \textit{et al.} \cite{Christian2012,Christian2012b} determined the ground state of $^{28}$F to be a resonance at 220(50) keV above the ground state of $^{27}$F,
is in good agreement with the USDA/USDB shell model predictions \cite{Brown2006},
suggesting that $pf$ shell intruder configurations play a minor role in the structure of the ground state of $^{28}$F,
even though calculations using the sdpf-u-mix interaction~\cite{Caurier2014} predict several low-lying negative parity states.

However, in 2020, Revel \textit{et al.} reported the first detailed spectroscopic study of $^{28}$F \cite{Revel2020}
and found the ground state of $^{28}$F to be a resonance at 199 keV with a negative parity.
The reconstructed $^{28}F+n$ momentum distribution following the neutron removal from $^{29}$F indicated that it arises mainly from the $p_{3/2}$ neutron shell,
with a $l=1$ content of about $80\%$, placing $^{28}$F inside the IOI.
In line with this finding, a subsequent study by Bagchi \textit{et al.} \cite{Bagchi2020} demonstrated the presence of a halo structure in the ground state of $^{29}$F
dominated by $l=1$ partial waves, and Gamow shell model calculations \cite{Michel2020} suggested the formation of a two-neutron halo in $^{31}$F.

Despite the experimental and theoretical studies mentioned previously,
our understanding of the structure of neutron-rich fluorine isotopes is still rather limited.
For that reason, the use of models with effective degrees of freedom, in which an intuitive interpretation of the mechanisms at play is possible,
can be helpful to better understand the structure of these systems.
Among the main factors that can impact the structure of neutron-rich fluorine isotopes are couplings to the continuum and nuclear deformation.

In this work, we study the single-particle structure of $^{28,29,31}$F by solving the Dirac equation
for one particle in a deformed Woods-Saxon potential obtained from relativistic mean-field (RMF) calculations \cite{Serot1986,Ring1996,Vretenar2005,Meng2006},
similarly to what was done for the relativistic deformed case in Ref.~\cite{Fang2017}.
Compared to Ref.~\cite{Fang2017}, to extract resonance parameters with more accuracy,
we calculate the continuum level density (CLD) \cite{Suzuki2005} using the Green's function (GF) method \cite{Economou2006}
obtained in the so-called complex-momentum representation (CMR) \cite{Li2016}, which is introduced below.
This approach will allow us to start from a reasonable mean-field description of $^{28,29,31}$F,
from which we can study the effect of quadrupole deformation while accounting for couplings to the continuum.

The RMF approach has been used to successfully describe masses and radii in stable nuclei \cite{Vretenar2005,Meng2006,Liang2015}
as well as to contribute to nuclear astrophysics problems \cite{Sun2008,Niu2009,Niu2011,Xu2013,Niu2013},
but it can also be extended for the explicit description of decaying resonances in exotic nuclei \cite{Yang2001,Zhang2004,Zhang2008,Guo2010}.
Here, as mentioned above, couplings to continuum states are included in the Dirac equation using the CMR,
which is a particular case of Berggren basis \cite{Berggren1968} built using spherical Bessel functions analytically continued in the fourth quadrant of the complex-momentum plane.
The idea is to represent the problem at hand in a basis including states having explicitly outgoing asymptotics,
so that its diagonalization provides eigenvectors whose asymptotics can be that of either a bound state, a scattering state, or a decaying resonance,
and its eigenvalues $\tilde{E}$ can be real $\tilde{E} = E$ or complex with $\tilde{E} = E -i\Gamma/2$ where $E$ is the energy position and $\Gamma$ the width of the resonance.

While the Berggren basis has been applied in various many-body calculations \cite{Michel2009,Rotureau2006,Hagen2007,Stoitsov2008,Wang2017,Hu2019},
the CMR specifically has been applied in the context of the RMF approach to study resonances in both spherical \cite{Li2016,Ding2018} and deformed \cite{Fang2017} nuclei,
and, like all Berggren basis expansion methods, has the advantage of describing both narrow and broad single-particle resonances naturally, providing the completeness of the basis is ensured.
Another way to probe the resonant single-particle structure of a nucleus is to look at the CLD mentioned previously.
This method offers an increased accuracy for resonances close to $E=0$ \cite{Luo2020,Shi2018} as compared to the CMR alone 
or approaches based on the complex-scaled Green's function method as in Ref.~\cite{Shi2015,Shi2016,Shi2017},
where a small dependence on the mathematical parameters of the method can remain.
In this work, all single-particle energies and widths are extracted from the CLD using the Green's function in the CMR.

We introduce the formalism of the RMF+CMR+GF approach and the numerical details in Section \ref{sec_formalism},
present the results in Section \ref{sec_results}, and give the conclusion in Section \ref{sec_conclusion}.


\section{Model and methods}
\label{sec_formalism}

The starting point of this work is the RMF framework in the Berggren basis.
This relativistic many-body method with nucleons as degrees of freedom allows us to generate effective spherical mean-field potentials for each isotope considered
using a single nucleon-nucleon interaction.
These spherical potentials are then used to fix the parameters of Woods-Saxon (WS) potentials where quadrupole deformation can be explicitly included,
and whose resonant spectra can be determined using the Dirac equation for deformed systems.
Here, we use the NL3 interaction \cite{Lalazissis1997}, which is given by the parametrization of an effective nonlinear Lagrangian density of relativistic mean field theory,
and which provides a good description of both nuclei at and away of the valley of $\beta$ stability.

For open-shell nuclei, pairing correlations are handled in the BCS approximation.
They are included using the constant gap approximation by occupation numbers of BCS-type \cite{Reinhard1986},
which is possible in this work because resonant states are well separated from the continuum.
Basically, it is assumed that pairing matrix elements are constant in the vicinity of the Fermi level \cite{Ring_book}.
When the resonances are accounted for, pairing correlations can be dealt with using the gap equation:

\begin{align}
& \sum_{i} \frac{\Omega_{i}}{\sqrt{\left(\varepsilon_{i}-\lambda\right)^{2}+\Delta^{2}}}
=\frac{2}{G}, \label{Diraceq13}
\end{align}
and the particle number equation:

\begin{align}
	& \sum_{i} \Omega_{i}\left[1-\frac{\varepsilon_{i}-\lambda}{\sqrt{\left(\varepsilon_{i}-\lambda\right)^{2}+\Delta^{2}}}\right] = N,
\end{align}
where $G$ is the pairing strength, $N$ is the particle number,
and $\Omega_{i}=j_{i}+\frac{1}{2}$ with $i$ for single-particle state.

In a second step, the spherical mean-field potentials obtained for each considered isotope and using the same interaction
are replaced by WS potentials adjusted as to reproduce the single-particle spectra of the mean-field potentials.
These WS potentials have the advantage of including quadrupole deformation explicitly (see below),
so they can be used in the Dirac equation for deformed systems.

The Dirac equation for a particle moving in a repulsive vector potential $V(\vec{r})$ and an attractive scalar potential $S(\vec{r})$ can be written as:

\begin{equation}
	[\vec{\alpha} \cdot \vec{p}+\beta(M+S(\vec{r}))+V(\vec{r})] \psi(\vec{r}) = \varepsilon \psi(\vec{r}),
	\label{Diraceq1}
\end{equation}
where $\vec{\alpha}$ and $\beta$ are Dirac matrices, $M$ and $\vec{p}$ are the nucleon mass and momentum, respectively.
$E=\varepsilon-M$ represents the single particle energy, and $\psi$ is the  wave function.
To introduce quadrupole deformation in the Dirac equation,
we write the vector and scalar potentials $V(\vec{r})$ and $S(\vec{r})$ as follows:

\begin{equation}
	\begin{array}{l}
		V(\vec{r})=V_{0} f(r)-\beta_{2} V_{0} k(r) Y_{20}(\cos\theta) \\
		S(\vec{r})=S_{0} f(r)-\beta_{2} S_{0} k(r) Y_{20}(\cos\theta)
	\end{array}
\label{eq_V_S}
\end{equation}
where $\beta_2$ is the quadrupole deformation parameter, and the radial functions $f(r)$ and $k(r)$ take WS forms:

\begin{equation}
	f(r) = \frac{1}{1+e^{\frac{r-R}{a}}},
	\label{eq_f}
\end{equation}
and $k(r) = r df(r)/dr$.
One note that the form of Eq.~\eqref{eq_V_S} is similar to the first order expansion in $\beta_2$ of a deformed WS potential,
and the usual spin-orbit coupling added in the Schr\"odinger equation is automatically included in the Dirac equation.
To obtain the resonant spectrum of the potentials in Eq.~\eqref{eq_V_S},
we include couplings to continuum states by expressing the Dirac equation in the complex-momentum representation:

\begin{equation}
	\int d \vec{k}^{\prime} \braket{ \vec{k} |H| \vec{k}^{\prime} } \psi(\vec{k}^{\prime}) = \varepsilon \psi(\vec{k}),
	\label{Diraceq2}
\end{equation}
where $H=\vec{\alpha} \cdot \vec{p}+\beta(M+S(\vec{r}))+V(\vec{r})$ is the Dirac Hamiltonian, $\psi(\vec{k})$ is the momentum wave function, and $\ket{\vec{k}}$  represents the wave function of a free particle with wave vector $\vec{k} = \vec{p}/\hbar$. For an axially deformed system, the parity $\pi$ and the third component of the total angular momentum $m_{j}$ are good quantum numbers.
Hence, $\psi(\vec{k})$ can be split into the radial and angular parts as:

\begin{equation}
	\psi(\vec{k})=\psi _{m_{j}}(\vec{k})=\sum_{lj}\left(\begin{array}{l}
		{f^{lj}(k) \phi_{lj m_{j}}\left(\Omega_{k}\right)} \\
		{g^{lj}(k) \phi_{\tilde{l}j m_{j}}\left(\Omega_{k}\right)}
	\end{array}\right).
	\label{Diraceq4}
\end{equation}
The angular part of the wave function is a two-component spinor,
represented as $\phi_{l j m_{j}}(\Omega_{k})=\sum\limits_{{{m}_{s}}} \braket{ lm\frac{1}{2}{{m}_{s}} | j{{m}_{j}} } Y_{lm(\Omega_{k})}\chi_{m_{s}}$.
The quantum number of the orbital angular momentum corresponding to the upper
(lower) component of Dirac spinor is denoted as $l(\tilde{l}$). The relationship between these two quantum numbers and the
total angular momentum quantum number $j$ reads $\tilde{l}=2 j-l$. Putting the wave function \eqref{Diraceq4} into the equation
\eqref{Diraceq2}, the radial Dirac equation becomes a set of coupled-channel equations:

\begin{widetext}
	\begin{align}
				M f^{lj}(k) - k g^{lj}(k) + \sum_{l' j'} \int k^{\prime 2} dk' V^+ (l',j',l,j,m_j,k,k',\theta) f^{l' j'}(k') = \varepsilon f^{lj}(k), \\
				-k f^{lj}(k) - M g^{lj}(k) + \sum_{l' j'} \int k^{\prime 2} dk' V^- (\tilde{l}',j',\tilde{l},j,m_j,k,k',\theta) g^{l' j'}(k') = \varepsilon g^{lj}(k),
			\label{Diraceq7}
	\end{align}
\end{widetext}
with:

\begin{align}
	& V^{+}(l',j',l,j,m_j,k,k',\theta) \nonumber \\
	& = {(-1)}^l i^{l+l'} \frac{2}{\pi} \int r^2 dr [V(r)+S(r)] j_l(kr) j_l'(k'r) \nonumber \\
	&\times \Theta(l,j,l',j',m_j, \theta),
	\label{eq_Vp}
\end{align}

and:
\begin{align}
	& V^- (\tilde{l}',j',\tilde{l},j,m_j,k,k',\theta) \nonumber \\
	& = {(-1)}^{\tilde{l}} i^{\tilde{l}+\tilde{l}'} \frac{2}{\pi} \int r^{2} dr [V(r)-S(r)] j_{\tilde{l}}(kr) j_{\tilde{l}'}(k'r) \nonumber \\
	&\times \Theta(\tilde{l},j,\tilde{l}',j',m_j, \theta),
	\label{eq_Vm}
\end{align}
where the angular factor is simply:

\begin{align}
	&\Theta(l,j,l',j',m_j, \theta) \nonumber \\
	&= \sum_{m_s} \braket{ l m | Y_{20}(\cos\theta) | l' m } \braket{ l m \frac{1}{2} m_s | j m_j } \braket{ l' m' \frac{1}{2} m_s | j' m_j },
	\label{eq_ang}
\end{align}
with $m = m_j - m_s$.

The coupled-channel equations can be seen in a matrix form and diagonalized,
and the solutions obtained be used to calculate the CLD.
One starts with the Green's function in momentum space defined as follows:

\begin{equation}
	G(E,\vec{k},\vec{k}') = \braket{ \vec{k} | \frac{1}{E-H} | \vec{k}'},
	\label{Diraceq9}
\end{equation}
and the extended completeness relation of Ref.~\cite{Myo2014}:

\begin{align}
	& \sum_{b}^{N_{b}} \ket{ \psi_{b}(\vec{k}) } \bra{ \tilde{\psi}_{b}(\vec{k}) } +\sum_{r}^{N_{\mathrm{r}}} \ket{ \psi_{r}(\vec{k}) } \bra{ \tilde{\psi}_{r}(\vec{k}) } \nonumber \\
	&\quad +\int \mathrm{d} E_{\mathrm{c}} \ket{ \psi_{c}(\vec{k}) } \bra{ \tilde{\psi}_{c}(\vec{k}) } = 1,
	\label{Diraceq10}
\end{align}
where $\psi_b(\vec{k})$, $\psi_r(\vec{k})$, and $\psi_c(\vec{k})$ denote the wave functions for the bound, resonant, and continuum (scattering) states.
The tilde in Eq.~\eqref{Diraceq10} denotes the Hermite conjugate.
Substituting Eq.~\eqref{Diraceq10} into Eq.~\eqref{Diraceq9}, the approximate density of states $\rho^N(E)$ with the basis number $N$ can be expressed as:

\begin{align}
	\rho^{N}(E) &= \sum_{b}^{N_{\mathrm{b}}} \delta\left(E-E_{\mathrm{b}}\right) \nonumber \\
	&+ \frac{1}{\pi} \sum_{r}^{N_{\mathrm{r}}} \frac{\Gamma_{\mathrm{r}} / 2}{\left(E-E_{\mathrm{r}}\right)^{2}+\Gamma_{\mathrm{r}}^{2} / 4} \nonumber \\
	&+ \frac{1}{\pi} \sum_{c}^{N-N_{\mathrm{b}}-N_{\mathrm{r}}} \frac{E_{\mathrm{c}}^{I}}{\left(E-E_{\mathrm{c}}^{R}\right)^{2}+E_{\mathrm{c}}^{I^{2}}},
\end{align}
where $E_b$, $E_r$, and $E_c$ are the eigenvalues for the bound states, resonances, and the continuum states, respectively,
$\Gamma_{\mathrm{r}}$ is the width of the resonance,
and $N_{b}$ and $N _{r}$ are the numbers of bound states and resonant states, respectively.
The CLD is then calculated as $\Delta \rho(E) = \rho^N(E) - \rho^N_0(E)$,
where $\rho^N_0(E)$ is the density of continuum states defined as:

\begin{equation}
	\rho_{0}^{N}(E) = \frac{1}{\pi} \sum_{c}^{N} \frac{E_{\mathrm{c}}^{0 I}}{\left(E-E_{\mathrm{c}}^{0 R}\right)^{2}+E_{\mathrm{c}}^{0 I^{2}}},
\end{equation}
and which corresponds to the level density of the asymptotic Hamiltonian $H_{0}$ for $r \to \infty$.


\section{Results}
\label{sec_results}

In this work, our goal is to explore the effect of quadrupole deformation and continuum couplings on the ground-state properties of $^{28,29,31}$F using the RMF theory with the CMR-GF method.
To this end, we adjust the parameters in the potentials defined in Eq.~\eqref{eq_V_S},
except for $\beta_2$ which is kept at zero, as to reproduce the single-particle levels of the self-consistent potentials obtained in RMF calculations using the NL3 interaction \cite{Lalazissis1997} in $^{28,29,31}$F.
One note that using other standard interactions such as the NL1, NL2, or PK1 give similar vector and scalar potentials.

With a similar fitting method as Ref.~\cite{Li2010}, we obtain the diffuseness of the potential $a=0.75$ fm, the radius $R=3.32$ fm,
and the depths of the vector and scalar potentials, $V_{0}=371$ MeV and $S_{0}=-439$ MeV for $^{28}$F and $V_{0}=337$ MeV and $S_{0}=-405$ MeV for $^{29,31}$F, respectively.
We note that some precision on the potential is lost in this procedure, but the overall properties of the mean-field are preserved.
Interestingly, for both sets of parameters $V_{0}+S_0 = -68$ MeV and the parameters obtained for $^{29}$F and $^{31}$F are identical.
In principle, different mean fields should emerge from the nucleon-nucleon interaction in different isotopes, which suggests that $^{29}$F and $^{31}$F have similar single-particle structures.
The resonance parameters can then be extracted by calculating the CLD while varying the quadrupole deformation.

\begin{figure}[htb]
	\includegraphics[width=1.0\linewidth]{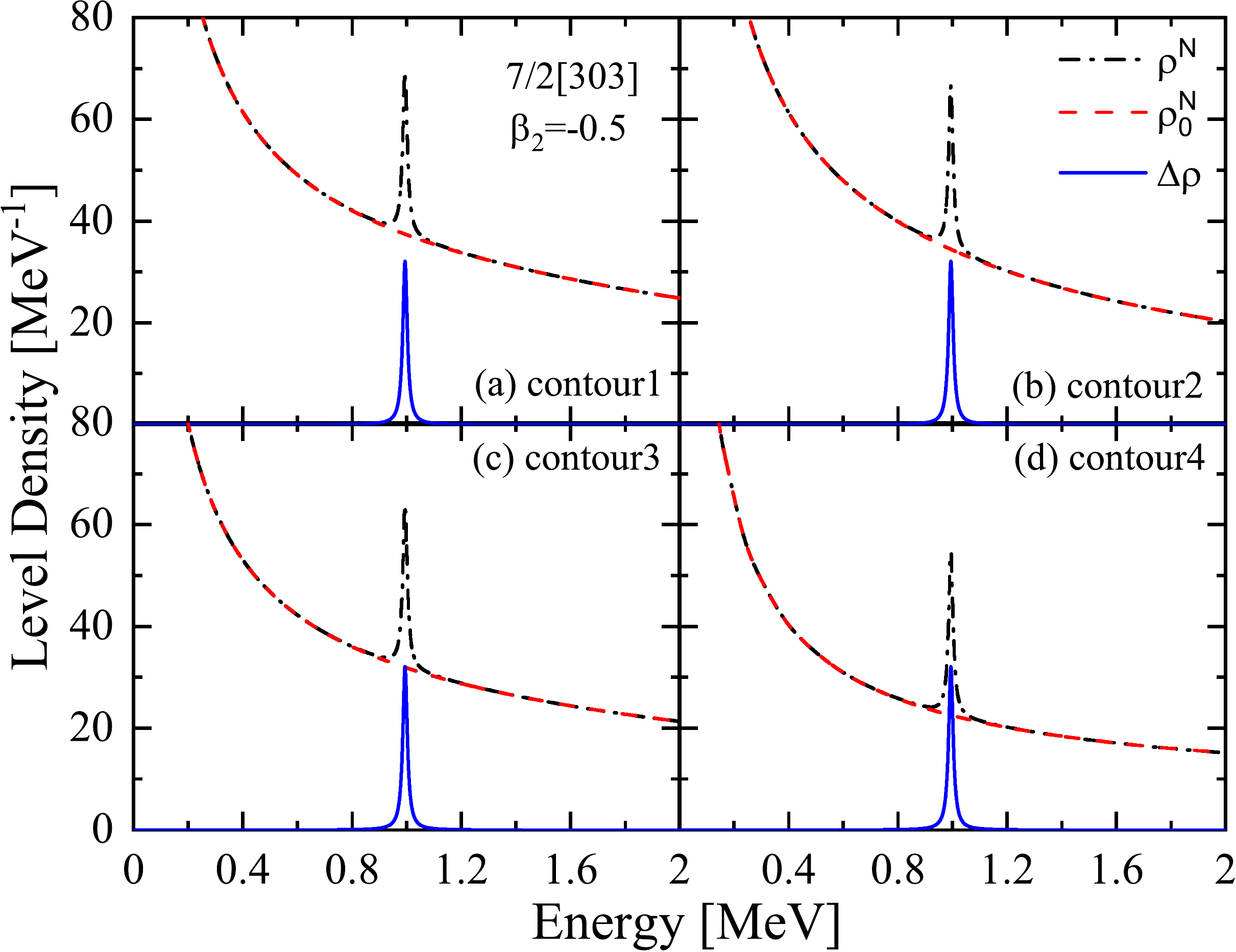}
	\caption{Density of states $\rho ^{N}(E )$, density of continuum states $\rho ^{N}_{0}(E )$, and CLD $\Delta \rho
	(E )$ for the 7/2[303] in $^{28}$F with $\beta _{2}=-0.5$ by the present calculations in four different contours for the momentum integration. These level densities are denoted with the black dash-dotted, red dashed, and blue solid lines, respectively.}
	\label{Fig1}
\end{figure}

We start by checking the completeness of the momentum expansion used to compute the CLD in the rest of the study.
In Fig.~\ref{Fig1}, we show the density of states $\rho ^{N}(E )$,
the density of continuum states $\rho ^{N}_{0}(E )$, and the CLD $\Delta\rho (E )$ for the resonant state 7/2[303] in $^{28}$F
with $\beta_{2}=-0.5$ for four different contours in the complex-momentum plane,
where 7/2[303] is labeled with the asymptotic quantum numbers $\Omega [ Nn_{z}\Lambda ]$.

These results were obtained using complex continua represented by four contours in the complex-momentum plane,
each made of three segments defined by the points $k_1 = 0.0$, $k_2$, $k_3 = 1.0$, and $k_\text{max} = 2.0$ (all in fm$^{-1}$),
and where $k_2 = (0.3 - i0.2)$, $(0.3-i0.25)$, $(0.2-i0.2)$, and $(0.8-i0.2)$ (all in fm$^{-1}$ for the contours 1, 2, 3, and 4, respectively.
Each segment was discretized using a Gauss-Legendre quadrature with a total of 120 scattering states on each contour,
which was enough to ensure proper convergence.

While both $\rho ^{N}(E )$ and $\rho ^{N}_{0}(E )$ change with the contours,
their difference, which gives the CLD $\Delta\rho (E )$ is invariant as it should be.
This will give us confidence to extract the energy and width of resonances of deformed potentials.
The corresponding position and width in a half height
of the resonance peak represent the energy and width of
resonant state, respectively.

We note in passing that for all other calculations, we will use the following enlarged contour as in Ref.~\cite{Fang2017} to avoid unnecessary readjustments:
$k_1 = 0.0$, $k_2 = (0.5-i0.5)$, $k_3 = 1.0$, and $k_\text{max} = 4.0$ (all in fm$^{-1}$).

\begin{figure}[htb]
	\includegraphics[width=1.0\linewidth]{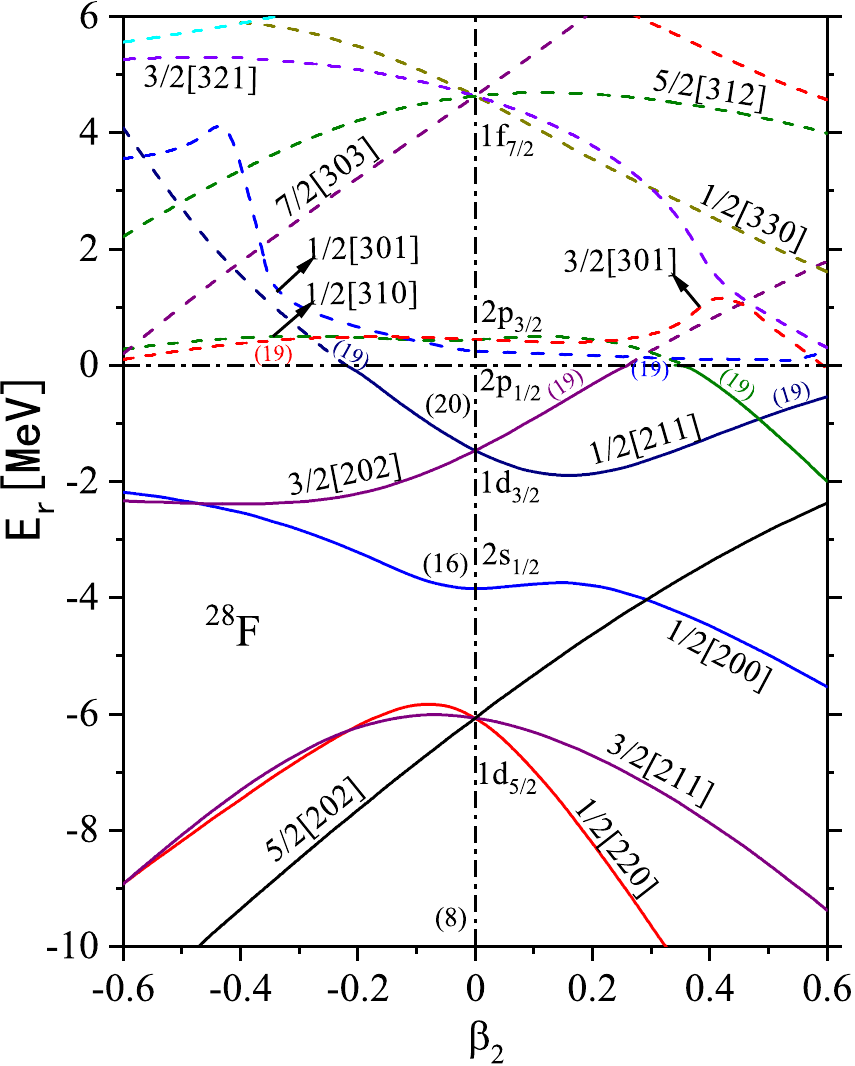}
	\caption{The neutron single-particle levels of $^{28}$F as a function of $\beta_{2}$. Every level is labeled with the Nilsson's asymptotic quantum numbers $\Omega [Nn_{z}\Lambda]$. The bound and resonant states are marked by solid and dashed lines, respectively.}
	\label{Fig2}
\end{figure}

Using the CMR-GF method, we calculate all the resonant neutron single-particle energies of $^{28}$F using the potential defined previously for $-0.6 \leq \beta_2 \leq 0.6$,
given by the energy positions extracted from the CLD $\Delta \rho(E )$,
and we label their trajectories using the quantum numbers $\Omega [ N n_{z} \Lambda ]$ corresponding to the so-called Nilsson orbitals.
The results are shown in Fig.~\ref{Fig2}.
Solid lines represent bound levels and dashed lines represent resonant levels (decaying resonances).
The corresponding spherical labels are marked in the position $\beta_{2}=0.0$.

We can see that, in the spherical case $\beta_2 = 0.0$,
the expected shell model order is respected up to the $1d_{3/2}$ shell,
which is about 1.5 MeV below the threshold.
However, the order of the $f$ and $p$ shells above the $1d_{3/2}$ shell differs from the expected shell model order.
The $2p_{1/2}$ and $2p_{3/2}$ shells lie close to the threshold, while the $1f_{7/2}$ shell is almost 5.0 MeV above it.

This result shows that the lowering of the $p$ shells,
which is a critical factor in obtaining a negative-parity ground state in $^{28}$F,
is not solely due to deformation.
Among the possible reasons for this lowering of the $p$ shells are the natural shell evolution due to tensor forces \cite{Otsuka2001,Otsuka2005},
and weak binding which tends to affect low-$\ell$ partial waves more strongly.

Nevertheless, according to the present result, the last neutron in $^{28}$F should occupy the $1d_{3/2}$ shell for $\beta_2 = 0.0$,
effectively giving a positive-parity ground state,
providing that the remaining neutrons form pairs coupled to $J^\pi = 0^+$ and that the last proton occupies a positive parity shell as expected from the shell model.
However, as such, this result would be in contradiction with the last experimental result on $^{28}$F \cite{Revel2020}.

At that point, we can discuss the role of deformation in the ground-state properties of $^{28}$F.
Indeed, when the spherical symmetry of the system is broken ($\beta _{2} \neq 0.0$),
the $1d_{3/2}$ and $2p_{3/2}$ shells split into their respective Nilsson orbitals ($1/2[211]$, $3/2[202]$) and ($1/2[310]$, $3/2[301]$),
while the $2p_{1/2}$ shell becomes the $1/2[301]$ orbital.
The energy positions of the Nilsson orbitals coming from the $p$ shells $1/2[310]$, $3/2[301]$, and $1/2[301]$
do not vary significantly in the typical range of deformation $-0.3 \leq \beta_2 \leq 0.3$,
but the energies of the $1/2[211]$ and $3/2[202]$ orbitals coming from the $1d_{3/2}$ shell
increase rapidly for oblate ($\beta_2 < 0$) and prolate ($\beta_2 > 0$) deformation, respectively.

This result indicates that even small amount of quadrupole deformation in either way,
can dramatically reduce the energy gap between positive and negative parity states in $^{28}$F.
The disappearing of the $N=20$ shell gap in IOI nuclei is, of course, largely due to quadrupole deformation,
and it is interesting to see hints of the same mechanism in $^{28}$F.

In fact, several shell model calculations \cite{Caurier2014} had already shown the emergence of deformation around $^{29}$F \cite{Utsuno2001},
providing that the $2p_{3/2}$ shell is included \cite{Poves1987}. The universal character of deformation in this region of the nuclear chart is simply a consequence of the Jahn-Teller effect \cite{Jahn1937,Reinhard1984}
due to the near-degeneracy of the $1d_{3/2}$ and $2p_{3/2}$ shells \cite{Hamamoto2007}.
The experimental finding of a negative-parity ground state in $^{28}$F thus suggests that some deformations might be present.

If one admits that some level of quadrupole deformation is present, then two questions remain.
Is the deformation prolate or oblate, and is the negative parity coming mostly from the $2p_{3/2}$ shell, as suggested by shell model calculations, or the $2p_{1/2}$ shell?

\begin{figure}[htb]
	\includegraphics[width=1.0\linewidth]{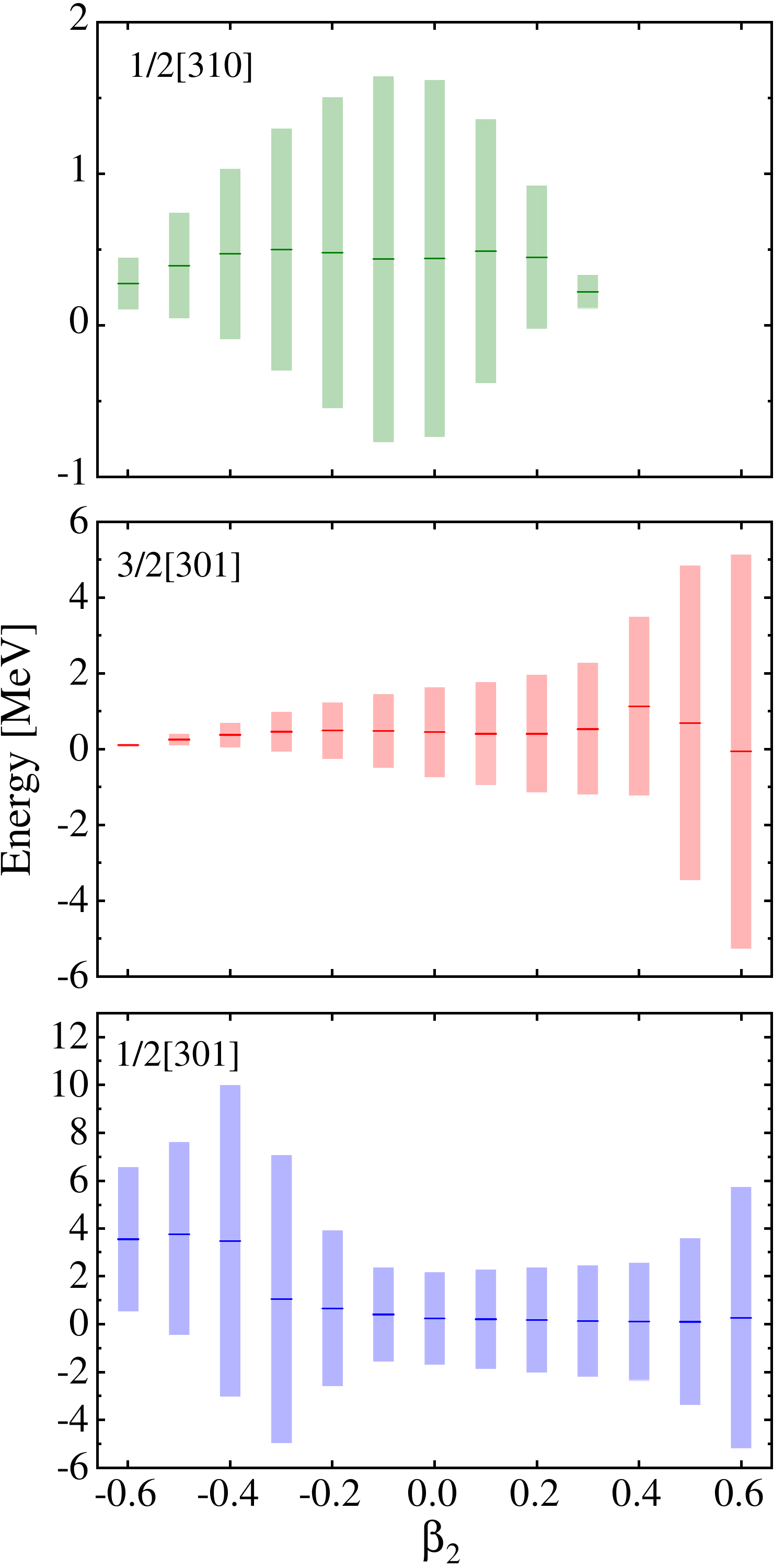}
	\caption{The single-particle energies and widths for 1/2[310],1/2[301],3/2[301] as a function of $\beta_2$. Heights of the rectangles represent the widths of this resonance.}
	\label{Fig3}
\end{figure}

The latter can be answered by looking at the widths of the Nilsson orbitals $1/2[310]$, $3/2[301]$, and $1/2[301]$ as a function of $\beta_2$ as shown in Fig.~\ref{Fig3}.
While the $1/2[301]$ Nilsson orbital ($2p_{1/2}$) is a broad resonance for all values of $\beta_2$, as shown in the bottom panel of Fig.~\ref{Fig3},
and the $3/2[301]$ (middle panel) becomes broader as $\beta_2$ increases, \textit{i.e.} for prolate deformation,
and the width of the $1/2[310]$ orbital (top panel) decreases rapidly with $\beta_2$ and vanishes (bound state) around $\beta_2=0.3$.
This suggests that the Nilsson orbitals coming from the $2p_{3/2}$ shell will dominate low-lying negative-parity states in $^{28}$F.

Whether the deformation should be prolate or oblate in $^{28}$F cannot be determined unambiguously from the qualitative nature of the present results.
However, the increased binding in heavier fluorine isotopes and the fact that $^{31}$F is bound,
a nucleus in which the $2p_{3/2}$ shell is presumably being filled,
suggests a lowering of this shell in $A>28$ fluorine isotopes.
This scenario favors a prolate deformation since, in the present model,
it leads to a drop of the $1/2[310]$ orbital below the threshold for positive values of $\beta_2$ in $^{28}$F.

Additional information can be gathered by assessing the weight of each single-particle state $\psi(\vec{k})$ in the Nilsson orbitals $\psi_{m_{j}}(\vec{k})$
by calculating the occupation probabilities $P^{m_{j}}$ as:

\begin{align}
	\operatorname{Re}(P^{m_{j}}) &= \int \widetilde{\psi}_{m_{j}}(\vec{k}) \psi_{m_{j}}(\vec{k}) \mathrm{d} \vec{k} \nonumber \\
	&= \int \sum_{lj} [f^{l j}(k) f^{l j}(k)+g^{l j}(k) g^{l j}(k)] k^{2} \mathrm{d} k.
\end{align}
We found that, for an oblate deformation of $\beta_2 = -0.3$,
the occupation probability of the $1f_{7/2}$ shell in the 1/2[310] and 3/2[301] Nilsson orbitals is limited to 3\% in both cases,
and while the 3/2[301] orbital is entirely dominated by the $2p_{3/2}$ shell (97\%),
the remaining occupation probabilities in the 1/2[310] orbital spilt almost evenly between the $2p_{3/2}$ (52\%) and the $2p_{1/2}$ (45\%) shells.

However, for a prolate deformation of $\beta_2=0.3$,
both the 1/2[310] and 3/2[301] orbitals are dominated by the $2p_{3/2}$ shell (77\% and 89\%, respectively)
and have a significant mixing with the $1f_{7/2}$ shell of about 8\% in both cases,
even though the 1/2[310] orbital in particular has a 15\% contribution form $2p_{1/2}$.
Overall, it appears that, compared to oblate deformation,
prolate deformation favors the occupation of the $2p_{3/2}$ shell as measured by the occupation probabilities for the 1/2[310] orbital.

\begin{figure}[htb]
	\includegraphics[width=1.0\linewidth]{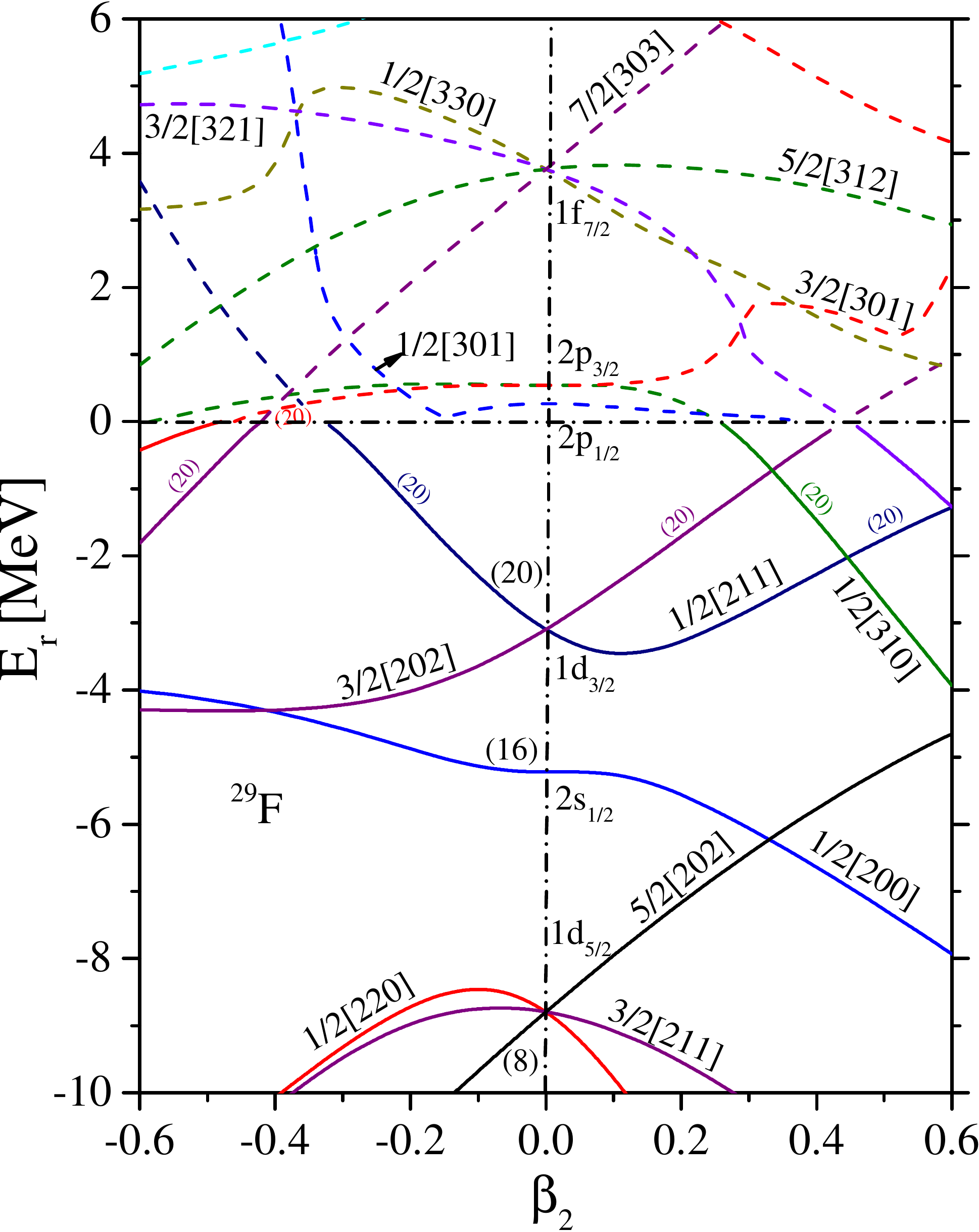}
	\caption{The neutron single-particle levels of $^{29}$F as a function of $\beta_{2}$. Every level is labeled with the Nilsson's asymptotic quantum numbers $\Omega [Nn_{z}\Lambda]$. The bound and resonant states are marked by solid and dashed lines, respectively.}
	\label{Fig4}
\end{figure}

To provide more perspective on the situation in neutron-rich fluorine isotopes in general,
we show the Nilsson diagrams for $^{29}$F in Fig.\ref{Fig4}.
One notes that a similar diagram is obtained in $^{31}$F (not shown).

The single-particle structure of $^{29}$F (and $^{31}$F) is similar to that of $^{28}$F in the spherical case,
with an inversion of $f$ and $p$ shells and $p$ shells close to the threshold,
but an important difference appears on the prolate side ($\beta_2 > 0$).
In $^{28}$F, the Nilsson orbitals $1/2[310]$, $3/2[301]$, and $1/2[301]$ associated with the shells $2p_{3/2}$ and $2p_{1/2}$, as well as $3/2[202]$
are all quasi-degenerate for a quadrupole deformation around $\beta_2 = 0.3$,
while in $^{29}$F (and $^{31}$F), the energy of the orbital $1/2[310]$ associated with $2p_{3/2}$ drops below the threshold
and becomes quasi-degenerate with the orbital $3/2[202]$ coming from $1d_{3/2}$ and which is more bound than in $^{28}$F.

Since it has been shown experimentally that the ground state of $^{29}$F presents a halo structure,
one can infer from this model that the quadrupole deformation in this isotope should be at least $\beta_2 = 0.3$.
One notes that in heavier isotopic chains (Ne, Na, Mg) the island of inversion typically starts at $N=20$
and is characterized by a sudden increase in quadrupole deformation.
For larger deformation the weight of $f$ waves increases and for that reason $\beta_2$ is unlikely to exceed significantly 0.45,
which is the point where the 1/2[211] and 1/2[310] orbitals cross and the 3/2[321] orbital coming from $1f_{7/2}$ becomes bound.

\begin{figure}[htb]
	\includegraphics[width=1.0\linewidth]{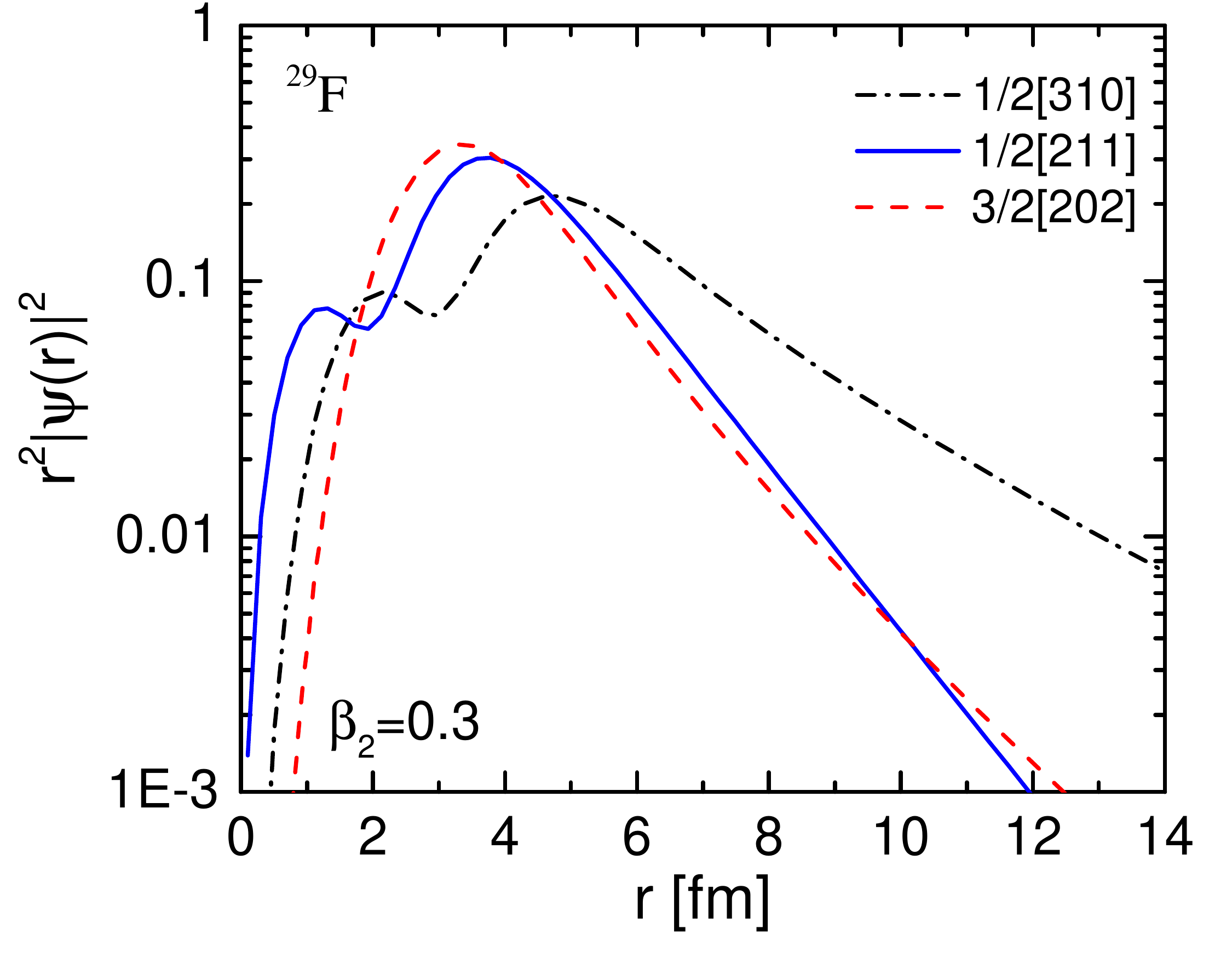}
	\caption{Radial density distributions for the single-particle states 1/2[310], 1/2[211], and 3/2[202] with $\beta_2=0.3$ in $^{29}$F.}
	\label{Fig6}
\end{figure}

\begin{figure}[htb]
	\includegraphics[width=1.0\linewidth]{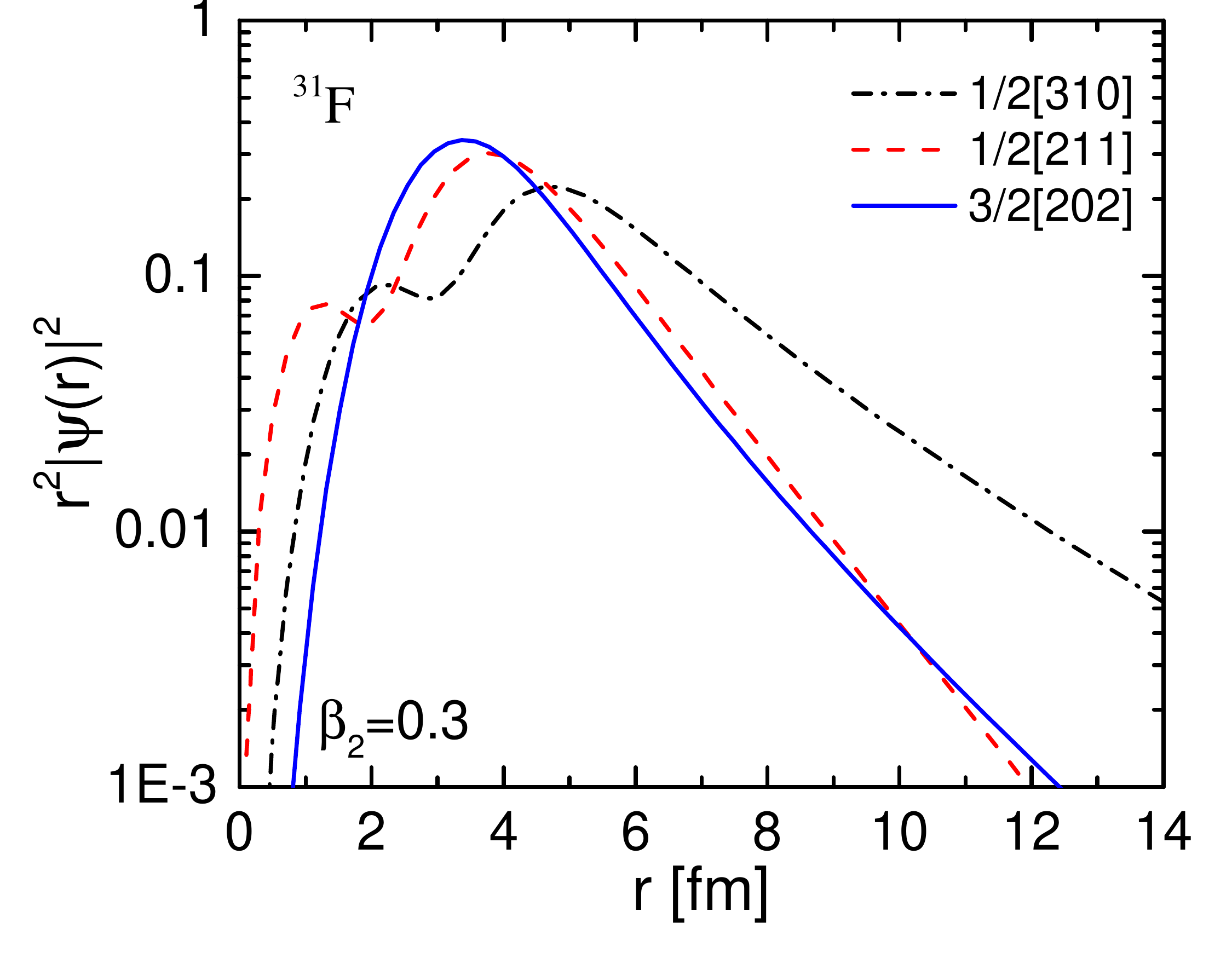}
	\caption{Radial density distributions for the single-particle states 1/2[310], 1/2[211], and 3/2[202] with $\beta_2=0.3$ in $^{31}$F.}
	\label{Fig7}
\end{figure}

To test whether or not a quadrupole deformation of $\beta_2 = 0.3$ is a reasonable guess in $^{29,31}$F,
in Figs.\ref{Fig6} and \ref{Fig7}, we show the radial densities associated with the bound Nilsson orbitals near the threshold in $^{29}$F and $^{31}$F, respectively. 
They are defined as:

\begin{equation}
	r^2 {| \Psi(r) |}^2 = r^2 \operatorname{Re} \sum_{lj} [f^{l j}(r) f^{l j}(r)+g^{l j}(r) g^{l j}(r)].
\end{equation}

In both cases,
the bound Nilsson orbital for $\beta_2 = 0.3$ associated with the $2p_{3/2}$ shell (1/2[310]) shows a characteristic halo tail at large distances
compared to those associated with the $1d_{3/2}$ shell (3/2[202], 1/2[211]).
As shown in Fig.\ref{Fig6}, the last two valence nucleons occupy the orbital 3/2[202] at $\beta_2=0.3$ and there is no halo phenomena. With the increase of $\beta_{2}$, the orbital 1/2[310] will gradually fall down, and the orbital 3/2[202] will rise. The last two valence nucleons will occupy the halo orbit 1/2[310], and the halo phenomena will appear in the larger deformation range. 
These results are in agreement with the observed halo in $^{29}$F \cite{Bagchi2020} and the predicted halo in $^{31}$F \cite{Revel2020}.

\begin{figure}[htb]
	\includegraphics[width=1.0\linewidth]{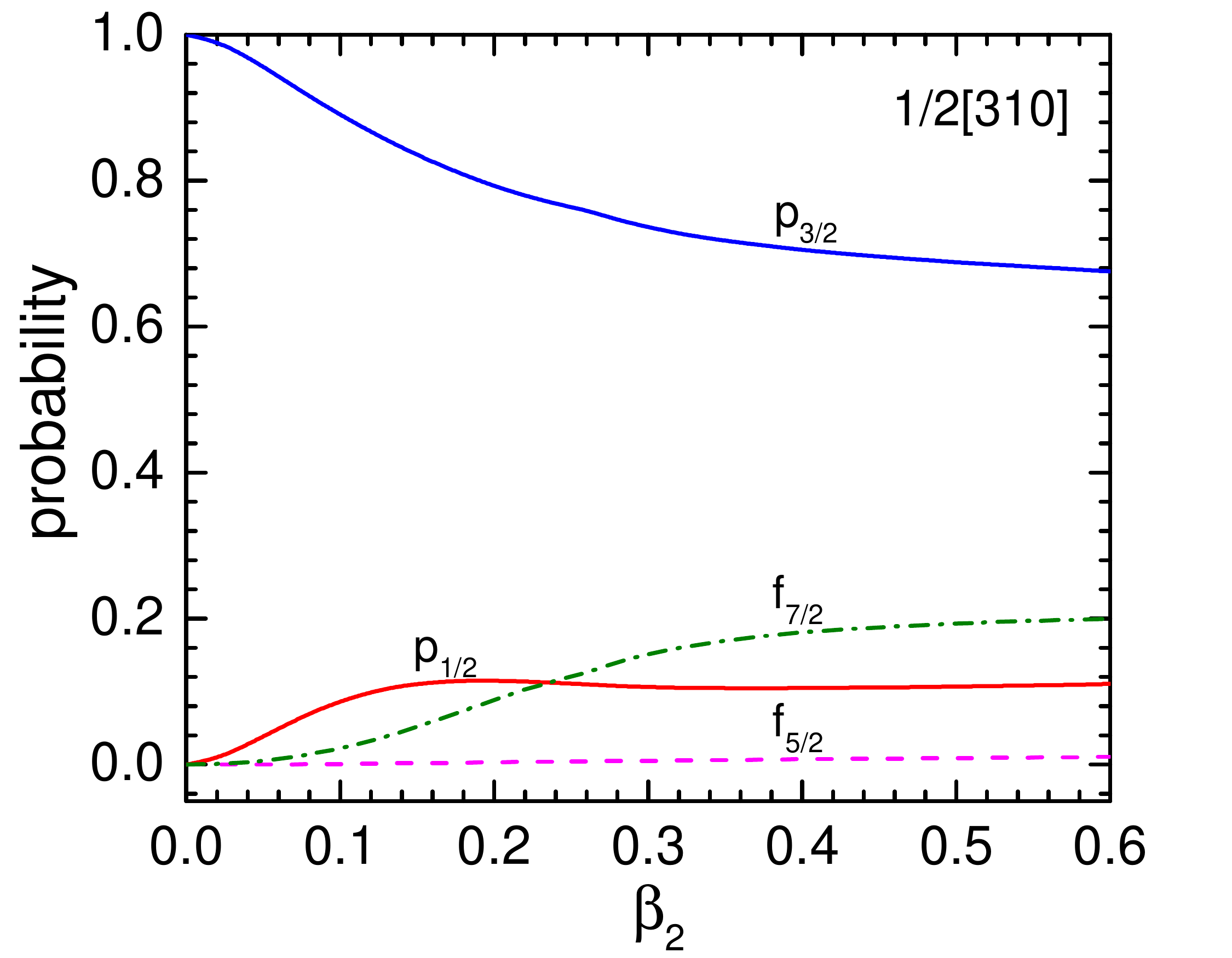}
	\caption{Occupation probabilities of major configurations as a function of $\beta_{2}$ for the single-particle state 1/2[310] in $^{29}$F.}
	\label{Fig8}
\end{figure}

Moreover, in $^{29}$F and for $\beta_2 > 0.3$, the $l=1$ content of the Nilsson state dominating the tail of the density at large distances (1/2[310])
is high ($\approx$ 80\%) and varies little with the quadrupole deformation as shown in Fig.~\ref{Fig8}.
The situation is nearly identical in $^{31}$F.

The results presented above show that it is possible to qualitatively explain the negative parity in $^{28}$F and the presence of halo structures in $^{29,31}$F within a single effective model
where both quadrupole deformation and couplings to the continuum are accounted for.
With all its limitations, the present approach allows to determine that some degree of prolate deformation is likely to be found in $^{28}$F,
and that quadrupole deformation must be between $\beta_2 = 0.3$ and 0.45 in both $^{29}$F and $^{31}$F.

Concerning neutron-rich fluorine isotopes in general,
the negative parity ground state in $^{28}$F is mostly due to couplings to the continuum but requires some level of deformation,
while the structure of heavier isotopes is dominated by an interplay between large quadrupole deformation starting at $N=20$ in $^{29}$F and couplings to the continuum.


\section{Summary}
\label{sec_conclusion}

In summary,
we have extended the CMR-GF method to the relativistic framework describing deformed nuclei,
and applied it to study the role of quadrupole deformation and continuum couplings on the single-particle structures of $^{28,29,31}$F.

We first extracted mean-field potentials for each considered isotope using the RMF method and for the same nucleon-nucleon interaction,
and used them to parametrize deformed WS potentials which served as input in the single-particle Dirac equation for deformed systems.
We then solved this equation in a typical range of quadrupole deformation
and extracted the energies and widths of the single-particle states in $^{28,29,31}$F
from the continuum level density $\Delta\rho(E)$ and demonstrated its reliability to study both narrow and broad resonances.

By building the Nilsson diagram for $^{28,29,31}$F including couplings to the continuum,
we found that without any deformation,
the $p$ shells already appear near the neutron-emission threshold, close to the $1d_{3/2}$ shell,
and any increase in quadrupole deformation (prolate or oblate),
leads to a rapid reduction of the gap between the $p$ and $d$ shells,
suggesting that some deformation is likely to exist in $^{28}$F and a quadrupole deformation of $0.3 \leq \beta_2 \leq 0.45$ is likely to be found in $^{29,31}$F.

In $^{28}$F, we showed that the analysis of the evolution of the widths of the relevant Nilsson orbitals
and of the occupation probabilities of various single-particle shells with deformation favors prolate deformation,
which is compatible with a negative parity ground state.
Additionally, in $^{29,31}$F, by looking at radial densities we demonstrated that the range of quadrupole deformation identified
gives typical halo distributions in $p$-wave dominated Nilsson orbitals as expected experimentally.

The experimental study of $A>29$ fluorine isotopes could gives important cues as to whether or not
quadrupole deformation develops with the number of neutrons as in other IOI nuclei,
and if so, how it affects the single-particle structure with increasing couplings to the continuum.

\begin{acknowledgments}
This work was partly supported by the National Natural Science Foundation of
China under Grants No.11935001 and No.11805004; the Key Research Foundation of Education Ministry of Anhui Province under Grant No.
KJ2018A0028, and the Natural Science Foundation of Anhui Province under Grants No. 2008085MA26.
This material is based upon work supported by the U.S. Department of Energy,
Office of Science, Office of Nuclear Physics, under the FRIB Theory Alliance award DE-SC0013617.
\end{acknowledgments}

\end{CJK*}

\end{document}